\documentclass[journal=jacsat,manuscript=article]{achemso}

\usepackage[version=3]{mhchem} %
\usepackage{soul}
\usepackage{color}
\usepackage{multicol}
\usepackage{subfigure}
\usepackage[table,xcdraw]{xcolor}
\usepackage{array}
\usepackage{hyperref}

\author{Nihang Fu}
\altaffiliation{Equal contribution}
\affiliation[First University]
{Department of Computer Science and Engineering, University of South Carolina, Columbia, SC USA}
\author{Jeffrey Hu}
\altaffiliation{Equal contribution}

\affiliation[First University]
{Department of Computer Science and Engineering, University of South Carolina, Columbia, SC USA}
\author{Ying Feng}
\affiliation[Second University]
{Hangzhou University of Electronic Science and Technology, Hangzhou China}

\author{Gregory Morrison}
\affiliation[First University]
{Department of Chemistry and Biochemistry, University of South Carolina, Columbia, SC USA}

\author{Hans-Conrad zur Loye}
\affiliation[First University]
{Department of Chemistry and Biochemistry, University of South Carolina, Columbia, SC USA}

\author{Jianjun Hu}
\email{jianjunh@cse.sc.edu}
\affiliation[First University]
{Department of Computer Science and Engineering, University of South Carolina, Columbia, SC USA}

\title[An \textsf{achemso} demo]
  {Composition based oxidation state prediction of materials using deep learning}
\abbreviations{IR,NMR,UV}

\begin{document}

\begin{tocentry}

Deep learning neural language model is used to learn the relationships of materials compositions and elemental oxidation states and exploit this deep dark knowledge for accurate prediction oxidation states for materials given only their composition.

\end{tocentry}

\begin{abstract}
  Oxidation states are the charges of atoms after their ionic approximation of their bonds, which have been widely used in charge-neutrality verification, crystal structure determination, and reaction estimation. Currently only heuristic rules exist for guessing the oxidation states of a given compound with many exceptions. Recent work has developed machine learning models based on heuristic structural features for predicting the oxidation states of metal ions. However, composition based oxidation state prediction still remains elusive so far, which is more important in new material discovery for which the structures are not even available. This work proposes a novel deep learning based BERT transformer language model BERTOS for predicting the oxidation states of all elements of inorganic compounds given only their chemical composition. Our model achieves 96.82\% accuracy for all-element oxidation states prediction benchmarked on the cleaned ICSD dataset and achieves 97.61\% accuracy for oxide materials. %
  The web app for BERTOS can be accessed at \url{http://www.materialsatlas.org/bertos}

\end{abstract}

\begin{singlespace}
\small

Oxidation states (OS) are the charges of atoms after their ionic approximation of their bonds, which are the fundamental attributes of elements that help to explain redox reactions, reactivity, chemical bonding, and chemical properties of different elements and compounds \cite{walsh2018oxidation,jorgensen2012oxidation}. In electrochemistry, oxidation states are used to represent relevant compounds and ions in Latimer and Frost diagrams, and they can also be used to calculate the charge neutrality of chemical compounds to screen potential hypothetical materials generated by computational design algorithms \cite{dan2020generative}. Oxidation states have also been used to study the complexes of transition metals \cite{karen2015oxidation}. In material science, oxidation states are useful in determining why one compound might be more suitable than another and have also been used in structure prediction for materials discovery by chemical analogy \cite{davies2018materials}. %

The chemist determine the atomic oxidation states of a compound with the help of many conditions such as extrapolation of a chemical bond's polarity (from electronegativity differences, dipole moments, or quantum-mechanical calculations of charges) or assigning electrons based on the number of electrons contributing to the molecular orbit. Commonly used rules for assigning oxidation states in chemical equations are as follows: the cation (positively charged ions which are metals) is written first, then the anion. The oxidation number of free elements is always 0, the oxidation of monatomic ions is the charge of the ion itself, the general oxidation number of hydrogen is +1, the general oxidation number of oxygen is -2 (-1 in compounds with elements with stronger electronegative than oxygen), the oxidation number of Group IA elements is +1, the oxidation number of Group IIA elements is +2, the oxidation number of Group VIIA elements is -1 (except when the element is combined with another element of higher electronegativity), the sum of oxidation numbers of atoms in neutral compounds is always 0, and the sum of oxidation numbers in polyatomic ions is equal to the charge of the ion. 
The challenge of assigning oxidation states lies in the fact that many elements may take different oxidation states based on their local chemical environment or bonding atoms, and most rules are only applicable to a few situations with different types of exceptions. To address this challenge, recently machine learning based algorithms have been proposed for oxidation states assignment for metal atoms in compounds \cite{jablonka2021using,amin2022predicting,shevchenko2022mining,reeves2019automated}.  
Jablonka et al.\cite{jablonka2021using} proposed the use of a machine learning model, trained with data from the Cambridge Structural Database (CSD), to automatically assign oxidation states to the metal ions in metal–organic frameworks. They found that charge-partitioning based computational techniques are unable to remove ambiguity, so instead, they proposed the use of collective knowledge from chemists to assign oxidation states instead of the rule-based deduction method based on formal counting rules. In their approach, they analyzed chemical names in the CSD for the oxidation states of metal centres, encoded the local chemical numerically, and trained a group of machine learning (ML) models. The models then made a predictions based "vote" between four base models to classify the oxidation states.
Amin et al.\cite{amin2022predicting} used supervised and unsupervised machine learning for predicting the oxidation states of Mn ions in the oxygen‑evolving complex of photosystem II by predicting the S-state of the X-ray, XFEL, and CryoEM structures. They trained a decision tree classifier and K-means clustering models using Mn compounds from the Cambridge Crystallographic Database.  The model uses two types of features: (1) The average bond length between the Mn and equatorial ligands and  (2) the axial ligands. The model was validated by using three different methods: predicting the Mn oxidation states in other protein structures with higher resolution, calculating the spin densities of the Mn in two of the mismatched structures of different S-states resolved by different groups, and validating the predictions by comparing them against the valence bond model.
In \cite{shevchenko2022mining}. Shevchenko et al. designed a random forest (RF) algorithm to predict the oxidation states of atoms and topological features of crystal structures using a predictive scheme for oxidation states in three stages: (1) selecting features and considering complex correlations between them, (2) training the RF model and choosing a set of hyper parameters, and (3) verifying the scheme with the testing samples by predicting oxidation states values and finding quality criteria. They selected features by evaluating relations between oxidation states and other descriptors with Pearson and Spearman correlations. The features were required to be strongly correlated with OxSt.
Automated oxidation-state assignment for metal sites in coordination complexes has also been studied in \cite{reeves2019automated}. In this work, Reeves et al. presented a automated workflow for oxidation-state assignment in transition-metal coordination complexes using the Cambridge Structural Database (CSD) Python API (application programming interface) scripts that complements the bond valence sum method for improved assignment confidence. 
Oxidation states of binary oxides from data analytics of the electronic structure was also investigated by Posysaev et al.\cite{posysaev2019oxidation}, in which a simple machine learning algorithm with linear regression is used to show that a correlation exists for the binary oxide systems. 

So far, all machine learning-based oxidation state prediction models are based on local structures of metal ions. There is no study on composition based ML models for OS prediction and there is no ML models for predicting the OS of non-metal elements with the structural information. In Pymatgen \cite{ong2013python}, there exists an oxidation state guessing function $oxide\_state\_guess$, which is based on an exhaustive enumeration algorithm and can give multiple possible charge-neutral oxidation state assignments. Currently, for composition based oxidation assignments, this can only be done using some heuristic rules by human experts or using the bond valence method, both of which have no automated software available for large-scale automated annotation of hypothetical candidate material compositions. In \cite{davies2018materials}, Davies et al. analyzed the preferences of cations and anions to take preferred oxidation states and summarized some rules for assigning oxidation states for compositions, which are then used to narrow down possible compositions for materials discovery. 

In this work, we proposes a novel deep learning based BERT transformer language model (BERTOS)  for composition-based oxidation state assignment for inorganic compounds. Our work is inspired by the recent success of deep learning in learning protein sequence patterns\cite{ferruz2022controllable}, predicting protein structures\cite{jumper2021highly}, predicting protein properties \cite{brandes2022proteinbert}, and even learning the structural patterns from protein sequences alone \cite{lin2022language}. Despite the difference of protein sequences and materials formulas, there have recently successfully developed deep learning based generative models for composition design of inorganic material compositions \cite{wei2022crystal,fu2022materials} using  generative adversarial networks \cite{dan2020generative} and transformer language models \cite{vaswani2017attention}, both of which are shown to be able to learn the chemical rules of materials and molecule compositions \cite{wei2022crystal,fu2022materials,wei2022probabilistic}. In these methods, the materials compositions and molecules are represented as sequences of element symbol sequences, which are fed to the transformer network trained using the self-supervised learning scheme. The trained language models were then used for generative design of new candidate material compositions or molecules. In this work, we map the composition-based oxidation state prediction problem as the token classification problem, in which the material formulas are converted into a sequence of element tokens sorted by their electronegativities and the output is the oxidation state of each atom. Our models are based on the BERT transformer language model, and have demonstrated great performance for oxidation states prediction. 

\section*{Results and Discussion}

The first step for oxidation state prediction modeling is to collect the datasets. Our raw dataset is obtained from the inorganic crystal structures database (ICSD)\cite{zagorac2019recent}. After cleaning and pre-processing as described in the Method section, we obtain four datasets including OS-ICSD (52,147 samples), OS-ICSD-CN (37,424 samples), OS-ICSD-oxide(35,886 samples), and OS-ICSD-CN-oxide(24,229 samples), each composed of a set of compositions along with OS assignments for all atoms. The sample statistics are shown in supplementary Table S1. By using a special data partition approach, we guarantee that there is no overlaps between any pair of training set and test set from these four datasets for performance evaluation.

\begin{figure*}[ht!] \centering    
   
\includegraphics[width=0.86\linewidth]{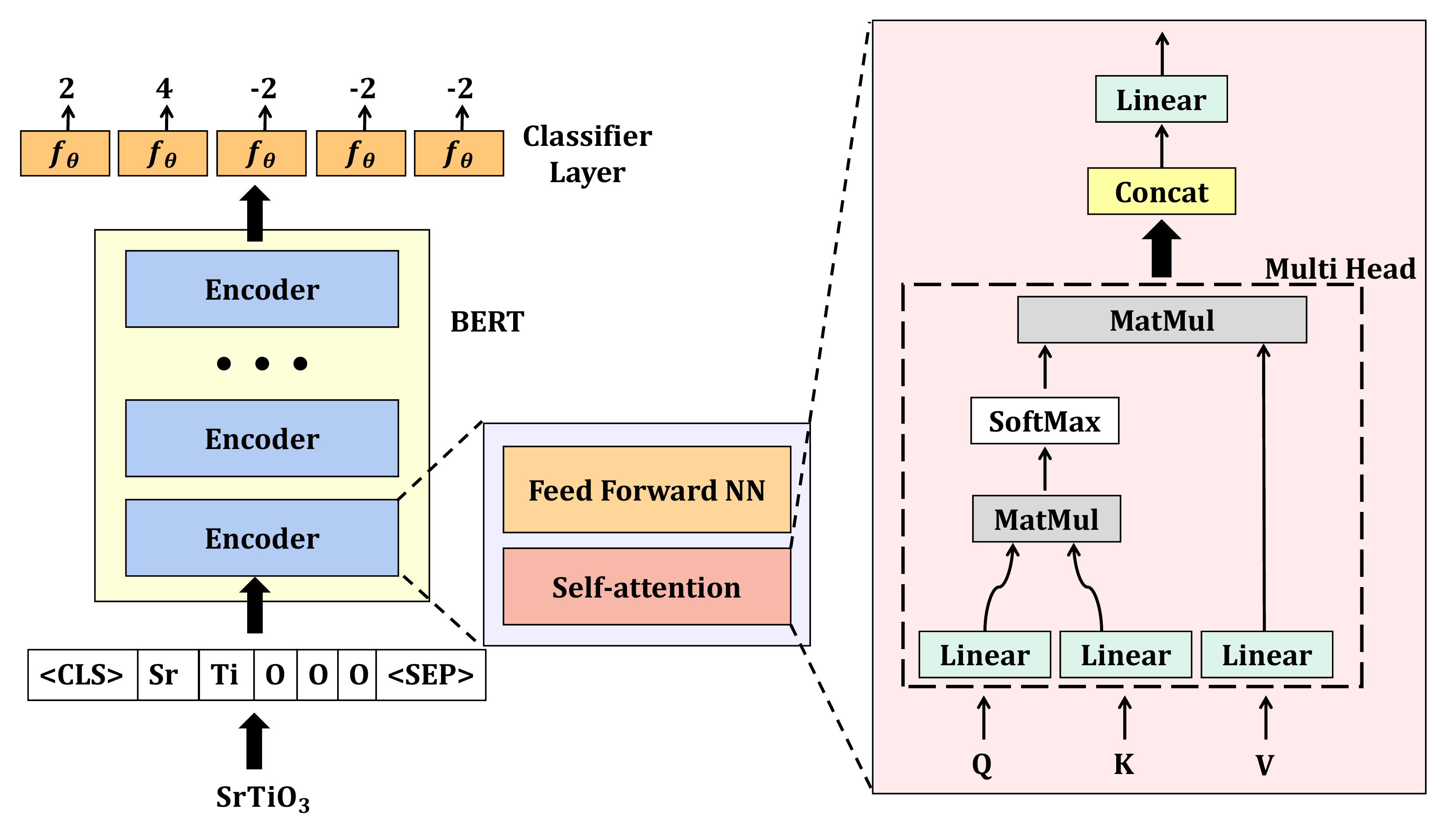}     
\caption{Architecture of BERTOS neural network model for oxidation state prediction. The overall framework (the left part) is composed of a BERT language model and a classifier layer $f_\theta$. Here, BERT consists of multiple stacked layers of encoders, and each encoder contains a self-attention network and a feed forward network (the middle part). The rightmost part is the detailed multi-head self-attention network, which uses the query, key, value mechanism to learn pair-wise correlations among tokens at different positions within a give window.}     
\label{fig:network}     
\end{figure*}

We formulate the composition-based oxidation state prediction problem as a token classification problem in natural language processing \cite{deshmukh2020deep}. We adopt the transformer language model \cite{wolf2020transformers} framework for token classification as shown in Figure\ref{fig:network}. It contains two modules including a BERT neural network for learning the token representation and a Dense network for the token classification. First we expand the regular material full formula/composition into a sequence of element symbol tokens. 
Using ternary compounds as an example, a typical material composition can be represented as $A_xB_yC_z$ where A/B/C are elements and x/y/z are the number of atoms of corresponding elements. The same rule applies to compounds with a different number of elements. If we only consider the cases where x/y/z are integers, we can expand the formula into $A_1 A_2...A_x B_1 B_2...B_y C_1 C_2..C_z$ where the elements are ordered by electronegativities. For example, $SrTiO_3$ can be expanded to $Sr\; Ti\;  O\;  O\;  O$, which becomes a regular sequence similar to a natural text sequence of words, a sequence of amino acids, or a SMILES representation of a molecule. With the expanded element sequence, our oxidation assignment problem for atoms becomes equivalent to the token classification in natural language understanding \cite{muthuraman2021data}.

The first module of our BERTOS framework is the Bidirectional Encoder Representations from Transformers (BERT) transformer language model. The BERT model is a transformer based neural networks originally designed for natural language processing tasks. The core of BERT is a transformer model with a variable number of encoder layers and self-attention heads (As shown in Figure \ref{fig:network}). BERT models can be trained on two tasks: masked language modeling and next token prediction to learn contextual embeddings for tokens in the training process. Here, we train the BERT model in a masked way. The raw representation of tokens are just one-hot encoding of elements without any additional elemental or structural properties. The goal of the BERT network is to learn the position-aware encoding of each token within the formula sequence context (such as the oxidation states based charge-neutrality constraints \cite{wei2022crystal}). The network can be used to generate the elemental embedding representation for each element symbol in the formula. 

The second module of our BERTOS framework is a shared regular feedforward fully connected neural network $f_{\theta}$ for mapping the BERT-generated element embedding into their corresponding oxidation states. $f_{\theta}$ receive the embeddings from the BERT network to predict the predicted labels for the input tokens. This network is trained together with the BERT network with the supervised signal of actual oxidation states of atoms within the formulas. The details of the network structure parameters and the training hyper-parameters are described in the supplementary file.

We first train a BEROS model using the cleaned OS-ICSD-CN training dataset with 31,827 samples with unique compositions and evaluate its performance over the OS-ICSD-CN test dataset with 3,724 unique compositions each labelled with ICSD-provided oxidation state labels. 

\paragraph{Overall performance:}We first check the atomic site-level accuracy $P_S$, which measures the percentage of correctly predicted oxidation states of all atoms in all compositions of the test set. Out of 190,468 atomic sites, our algorithm achieves 96.27\% accuracy (all elements are counted), which is close to the 98.1\% accuracy evaluated for only 994 metal atom oxidation states in 532 ionic and coordination compounds using structure based features and Random forest classifier. Our model even beats the structure based ML models for the Mn ion oxidation state prediction \cite{amin2022predicting} with accuracy of 94\% and 95\% for Gaussian Naive Bayes Classifier and decision tree classifiers tested only on Mn compounds. This result is surprising as our model only uses composition as input information. We then calculate $P_C$, the percentage of test materials that have all their atomic sites with correctly predicted oxidation states. The $P_C$ accuracy reaches 87.76\%. In contrast, when we use the Pymatgen's $oxid\_state\_guess$ function, only 4.49\% of the test samples can be assigned definite oxidation states. We further calculate the compound-level average site accuracy $P_{CASA}$, which reaches 97.16\%.

\begin{figure*}[ht] 
\centering    
\subfigure[] { 
\label{fig:nonmetalaccuracy}     
\includegraphics[width=0.45\linewidth]{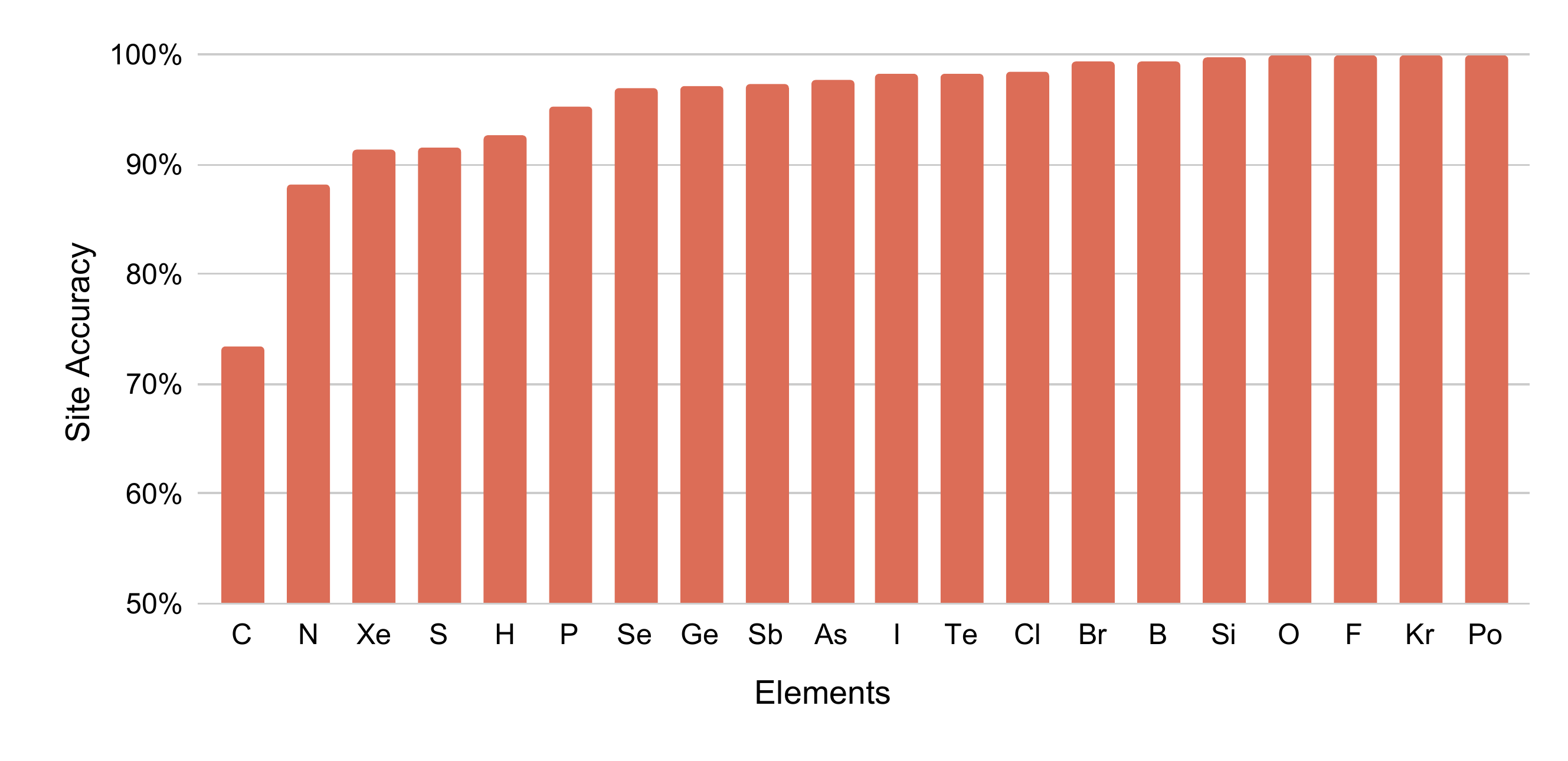}     
}    

\subfigure[] {
 \label{fig:metalaccuracy}     
\includegraphics[width=1\linewidth]{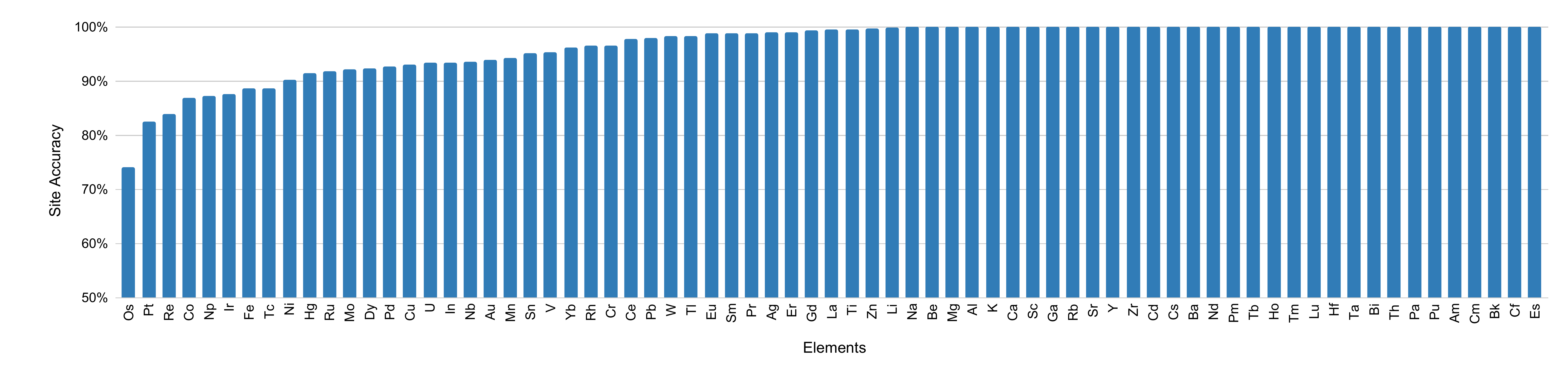}  
}     

\caption{(a) Site OS prediction accuracy for nonmetal elements. (b) Site OS prediction accuracy for metal elements.}     
\label{fig}     
\end{figure*}

\paragraph{Element family performance:} Since different elements take different set of possible oxidation states including those common OS ones, it is interesting to check our model performance over different element families.  
We first calculate the overall accuracy $P_e$ of all metal element sites, which reaches 97.12\%, which is slightly lower than 98.1\%, the accuracy of the structure-based machine learning model as reported in \cite{shevchenko2022mining}. This unexpected small gap to the structure-based model performance makes our model to be very useful in screening hypothetical materials compositions designed by deep generative models \cite{dan2020generative}. We then check the performance for individual element OS prediction. First we find that for all elements in the alkali metals and alkaline earth metals except Li (99.93\%), K (99.96\%), Fr and Ra (no data), our model achieves 100\% site accuracy, reflecting that our model learns to assign oxidation states to these strongly metal elements with high confidence, which is consistent with chemistry knowledge for which it is easier to assign OS. Next, we check the elements with low prediction accuracy on the left part of Figure\ref{fig:metalaccuracy}. Out of the elements with less than 90\% accuracy, a majority of them are transition metals including Os (74.07\%), Pt (82.54\%), Re (83.90\%), Co (86.89\%), Np (87.30\%), Ir (87.50\%), Fe (88.63\%), Tc (88.68\%). The relatively low performance over these transition metal elements reflects our model's learned chemistry is consistent with known literature. The other three elements with lower performance are from the Ac family including Fm(no data), Md (no data), No (no data), Lr (no data), and Np(87.30\%). We also find in general our model has higher OS prediction performance for the post-transition metal elements compared to transition metals. These elements include %
Sn(95.12\%), Pb(98.00\%), Ti(99.60\%), Ga(100\%), Bi(100\%), and Al(100\%).

In addition, we calculate the overall accuracy for all nonmetal element sites (including the metalloid elements), which reaches 96.05\% out of 150,778 sites, and the performance figure is shown in Figure \ref{fig:nonmetalaccuracy}. We then check the nonmetal elements with high performance, and find the OS prediction accuracy of the following elements can hit more than 99\%: Br (99.30\%), B (99.36\%), Si (99.76\%), O (99.90\%), F (99.95\%), Kr (100\%). There are only two non-mental elements, C (77.38\%) and N (88.20\%) with site accuracy less than 90\%, which is easier to explain because C and N have more common oxidation states than others, which makes it difficult for our deep learning models to learn the OS assignment rules. Other than that, the rest of nonmetal elements can maintain high site accuracy between 90\% and 99\%.

\begin{figure*}[ht]
  \centering
  \includegraphics[width=0.8\linewidth]{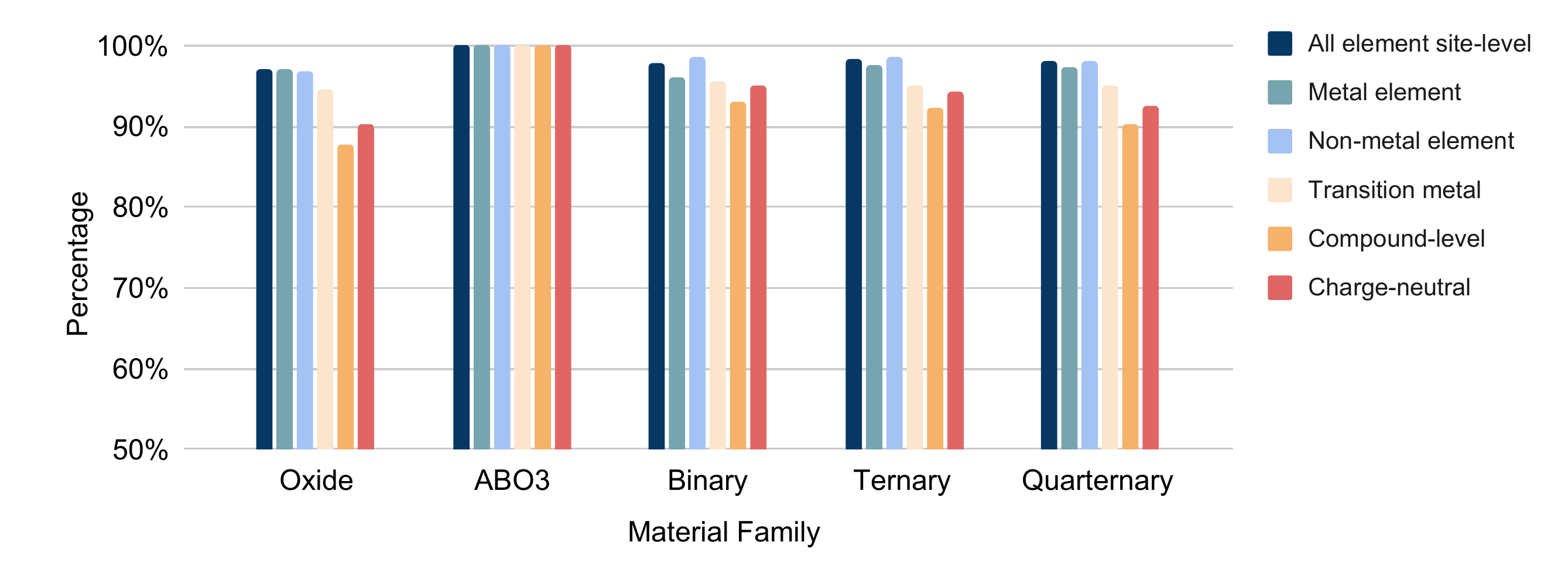}
  \caption{OS prediction performance of BERTOS over different material families. Except the compound level and charge-neutrality accuracy, all other performance criteria are at the site level.}
  \label{fig:mfamily}
\end{figure*}

\paragraph{Material family performance:}
We further examine how the OS prediction performance on a specific material family varies. We calculate the performances of our model over oxides, ABO3, binary, ternary and quaternary compounds with the results shown in Figure \ref{fig:mfamily}. First we check the OS performance over the binary, ternary, and quaternary materials. Over all element types, BERTOS achieves high accuracy, ranging from 97.77\% for binary to 98.03\% for quaternary compounds. Comparing the metal and non-metal elements, the non-metal elements tend to have higher prediction accuracy across all three families. For metal elements, the performance on ternary and quarternary families is higher than that of the binary compounds. We further find that the OS prediction performance on transition metals is lower than that on non-metal and the overall metal elements, which is consistent with chemical knowledge. At the compound level, all three families have accuracy greater than 87\%. We also check the charge-neutrality accuracy which measures what percentage of compounds that are charge-neutral as annotated in ICSD are still charge-neutral as predicted by BERTOS. Our algorithm achieves more than 92\% for all three families. 

We further check if BERTOS can achieve higher performance for more specialized materials families. We check the oxide materials and find that BERTOS achieves slightly higher performance than over binary materials for all elements, metal and non-metal elements. But the compound and charge-neutrality accuracy are not as good as those over ternary and quaternary ones. However, when we check the OS prediction performance of BERTOS on the $ABO_3$ materials, we find that all the performances are much higher than all other materials families, which demonstrates that the OS prediction performance can vary by material families with some of them are easier to predict as there are stronger OS rules there. 

\begin{figure*}[ht!]
  \centering
  \subfigure[non-metal elements]{
  \includegraphics[width=0.45\linewidth]{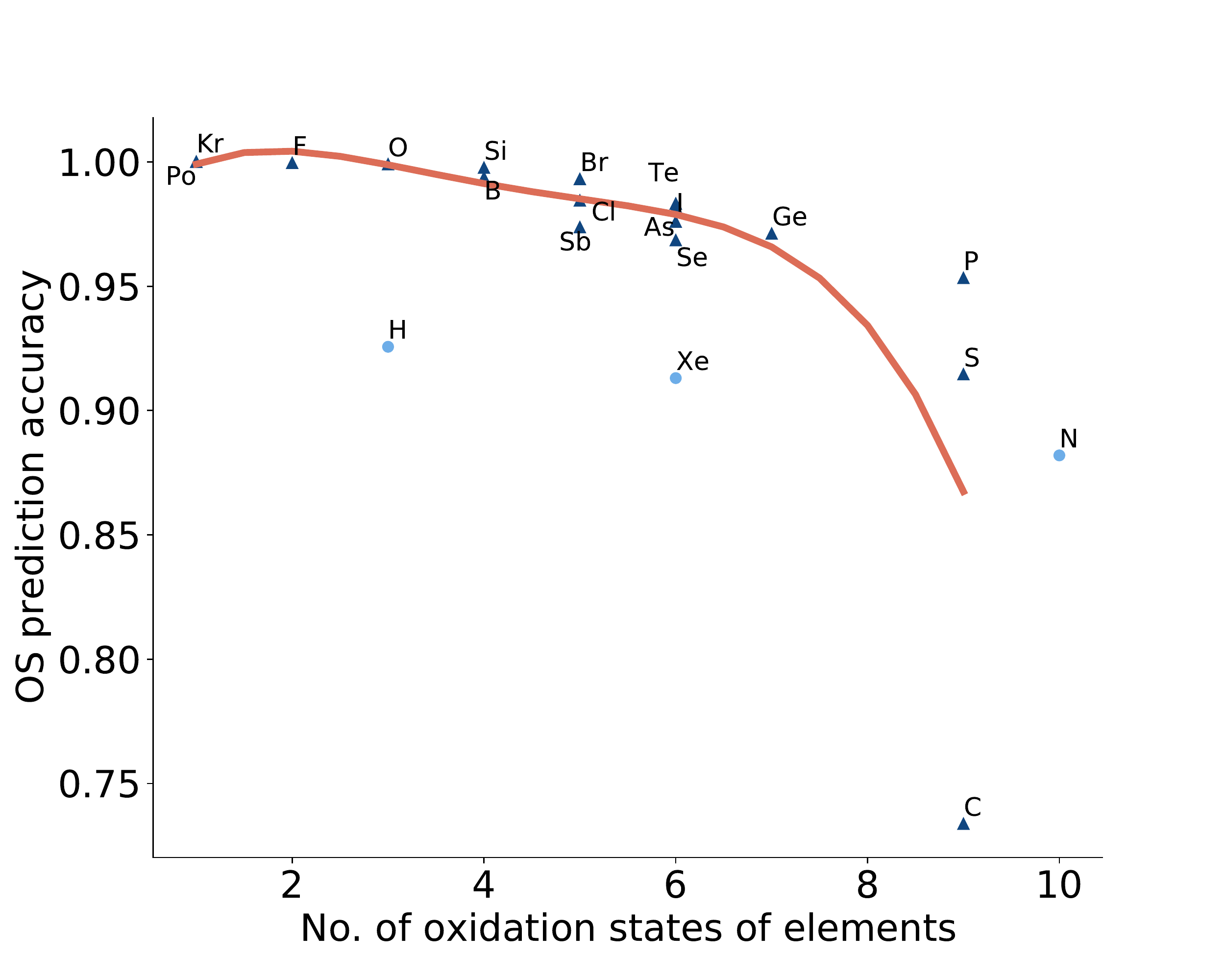}
  
  }
  \subfigure[metal elements]{
  \includegraphics[width=0.45\linewidth]{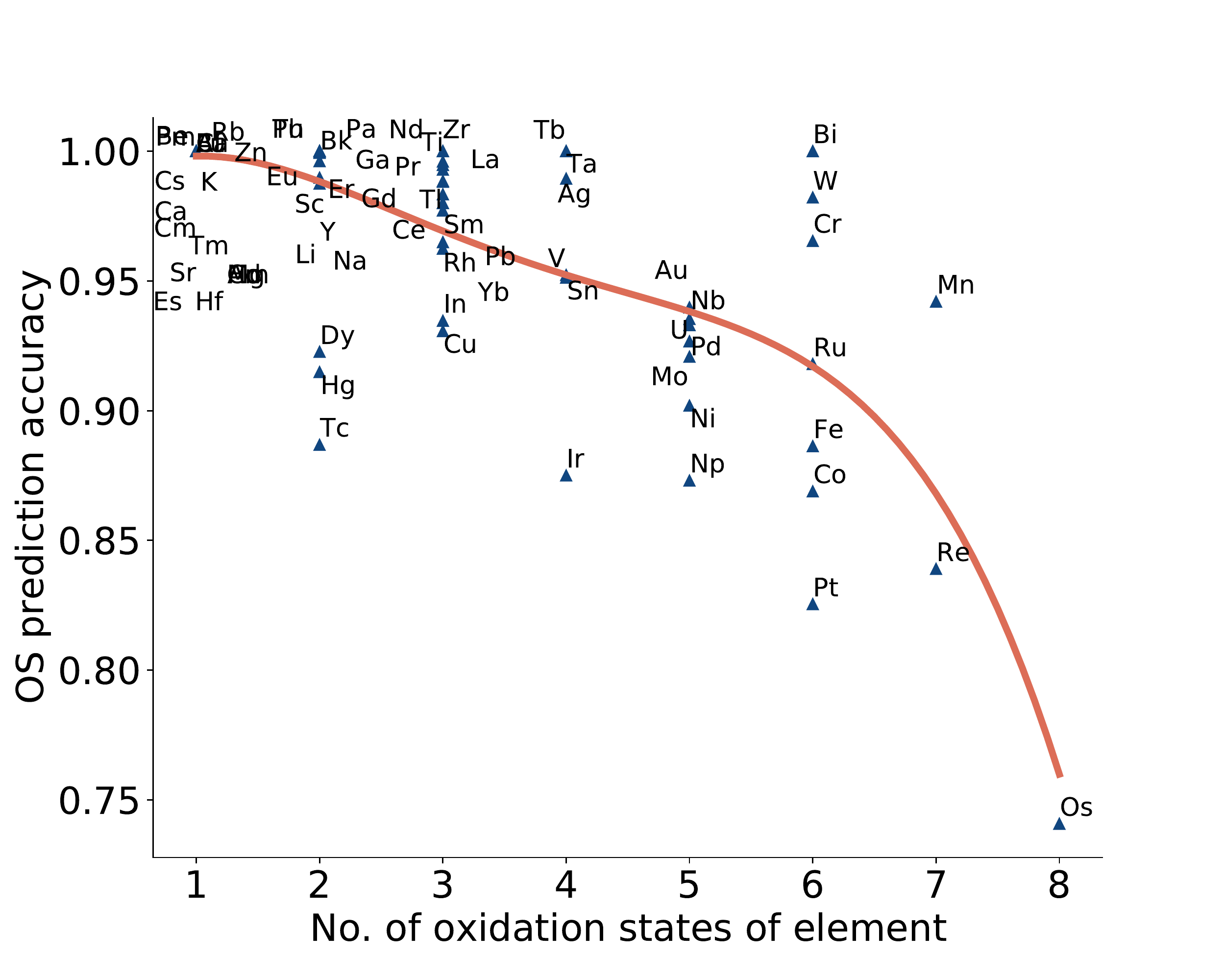}
  }  

  \caption{OS prediction performance versus no. of oxidation states. (a) non-metal elements.   The elements, H, Xe, and N are apparent outliers which have lower OS prediction performance compared to those elements with similar
numbers of accessible OS. (b) metal elements. For metal elements, while there is a pattern that the more the number of accessible oxidation states an element has, the more difficult it is to predict its oxidation states, there are many exceptions. The elements Tc, Ir, Np, Pt are apparent outliers which have lower OS prediction performance compared to those elements with similar numbers of accessible OS. Bi, W, Cr, and Mn instead have higher OS prediction accuracy compared to others with similar number of accessible OS.}
  \label{fig:accuracy_non_metal_osno}
\end{figure*}

We further check how the number of oxidation states of elements affect its OS prediction performance. Our common wisdom is that the fewer the oxidation states an element may have, the easier it is to predict its OS. Figure \ref{fig:accuracy_non_metal_osno} (a) shows how the non-metal element OS prediction performance of our model varies as regard to the number of possible oxidation states of a given element. We find that except the Hydrogen (H) and Xenon (Xe) element, there is a clear trend that when the element's OS set grows larger, it becomes more difficult to predict its OS. However, for the set of 10 elements more than 9 possible OS (C, P, S, N), the prediction accuracy changed from 73.38\% to 95.33\%, correctly reflecting their different degrees of chemical activities. We also checked the OS number versus prediction performance for metal elements as shown in Figure \ref{fig:accuracy_non_metal_osno}(b). We find that the correlation is much weaker compared to non-metal elements. While there is a trend that elements with more OS are more difficult to predict their OS, the variation is much higher: even elements with only five OS such as Th only achieves an accuracy of 85.45\% while the Mo element achieves 97.66\% despite it has 10 possible oxidation states.

\begin{figure*}[ht!]
  \centering
  \includegraphics[width=0.8\linewidth]{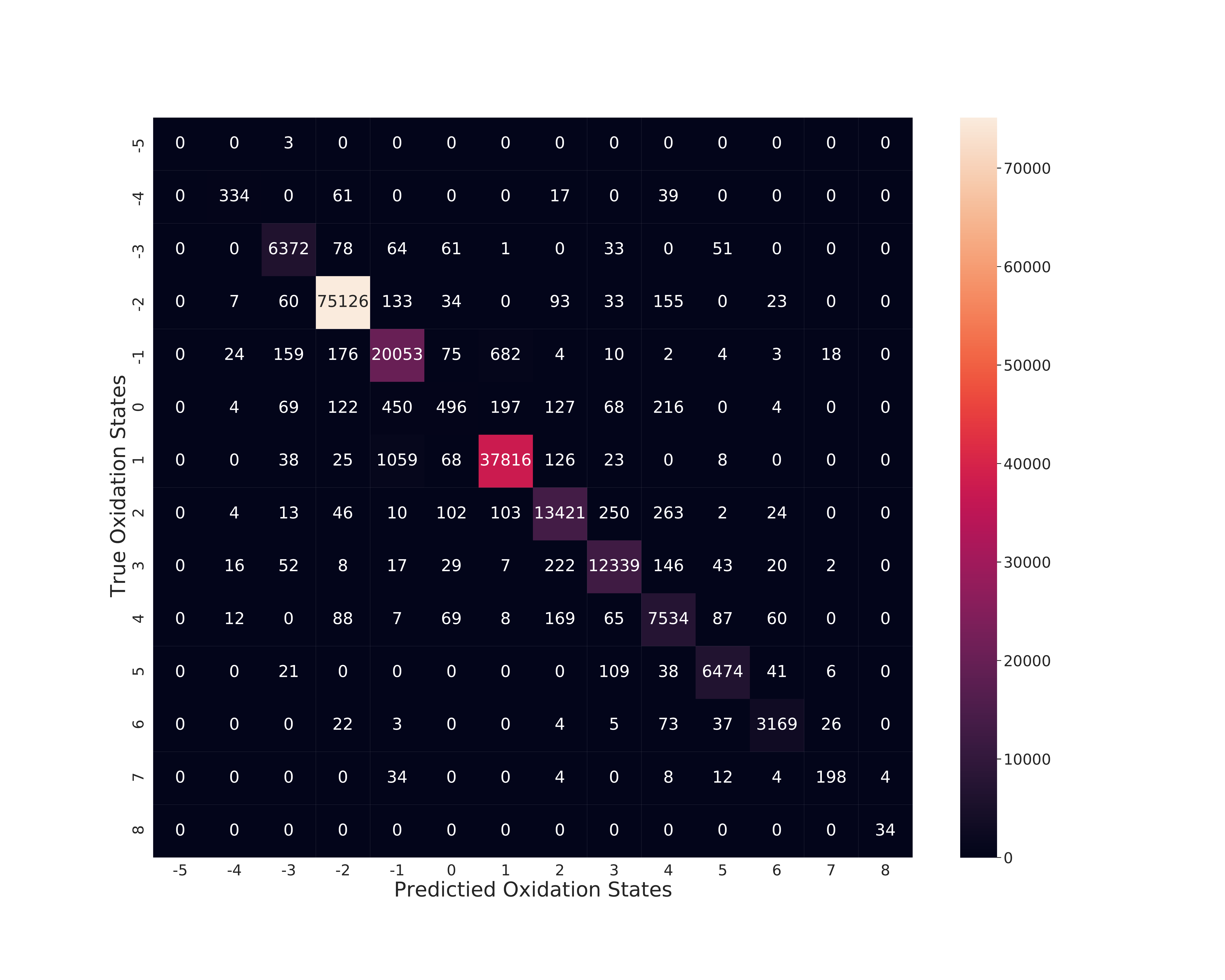}
  \caption{Confusion matrix of OS prediction evaluated on the OS-ICSD-CN test set. The x-axis represents predicted oxidation states, and the y-axis represents true oxidation states. The color scale on the right shows the relationship between the number of samples distribution and the shade of the color, which is the lighter the color, the more samples there are. Then, the numbers in the diagonal mean the number of the correct predictions.}
  \label{fig:confusion_icsdcn}
\end{figure*}

To explore the performance of OS predictions on each oxidation state (from -5 to +8), we plot the confusion matrix of the OS predictions for the OS-ICSD-CN test set (Figure \ref{fig:confusion_icsdcn}), and the overall accuracy of the predictions can hit 96.27\%. According to the shade of color, we find the oxidation states of most atomic sites concentrate between -3 and +6, and only a few atoms have OS distributed among -5, -4, 0, +7, and +8. Of all the OS, -2 has a total of 75,126 sites, which is the largest one with BERTOS prediction accuracy as high as 99.29\%, the highest performance among all the OS. One of the possible reason is that there are a lot of oxides or sulfides in our datasets, in which the oxygen and sulfur elements are usually take their common oxidation state of -2. We then check the OS prediction performance of BERTOS over other OS, and find with the exception of -5 (3 samples, 0\%),  -4 (451 samples, 74.06\%), 0 (1753 samples, 28.29\%), and +7 (264 samples, 75\%), the prediction accuracy of all other OS is more than 94\%. This is an interesting and reasonable since all of the OS with bad accuracy scores including -4, 0, and +7 are the ones with few occurrences. In addition to showing the overall OS prediction performance with regard to the atom sites, we also divide the atomic sites of the test set into metal and nonmetal ones and show their confusion matrices  (see Figure S2 %
and Figure S3 %
in the supplementary file). The overall accuracy of the metal atomic site OS reaches 97.12\%, and the nonmetal ones reaching 96.05\%. The most accurate OS for metal sites are +1, +2, +3 while the most accurate OS for non-metal sites are -2, -1, +1, +4.

\begin{figure*}[ht]
  \centering
  \includegraphics[width=0.8\linewidth]{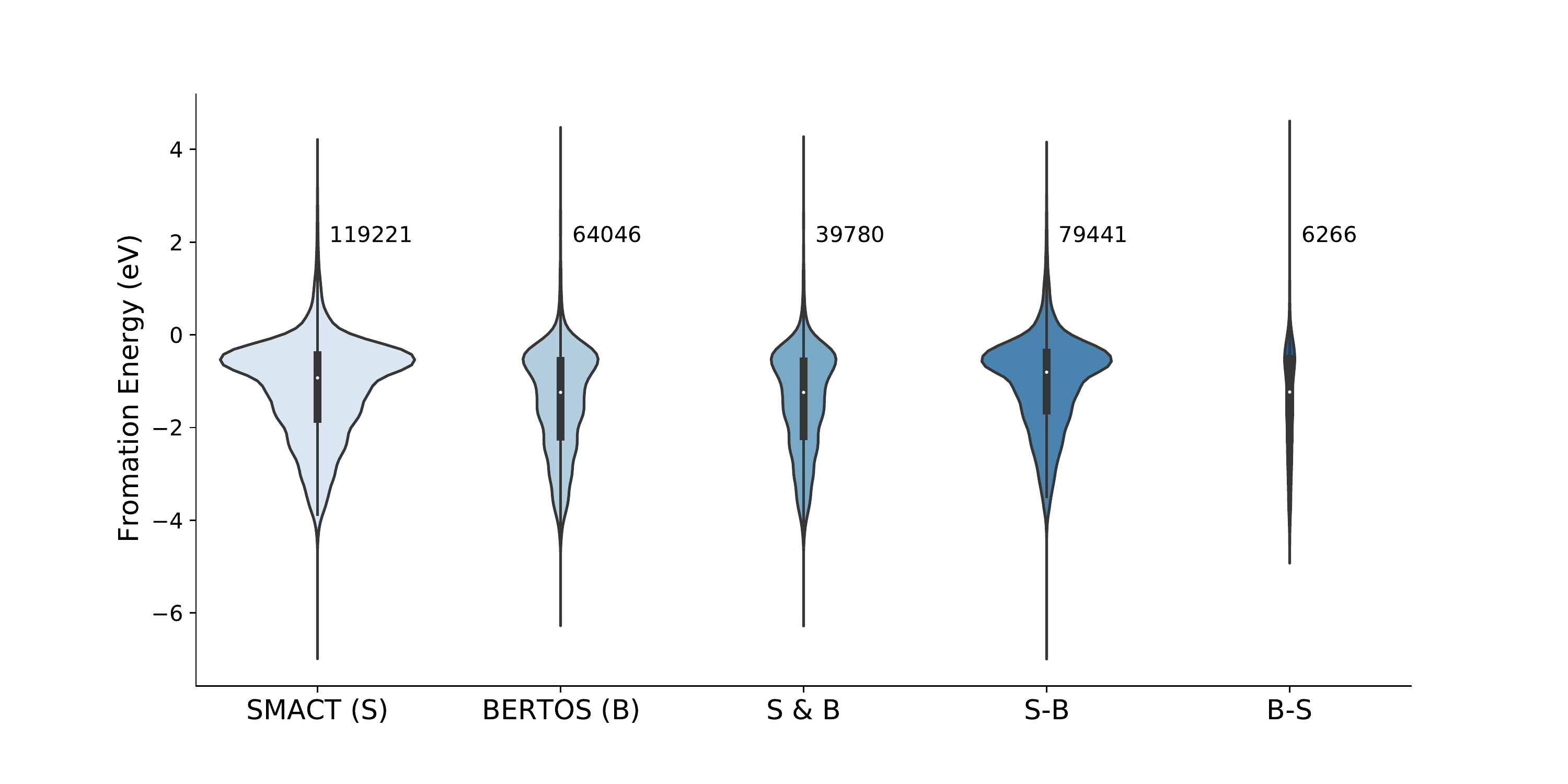}
  \caption{Comparison of formation energy distribution of binary and ternary candidate materials filtered by our BERTOS model trained on the OS-ICSD-CN dataset and by the enumerative charge-neutrality check algorithm of SMACT. S is the compositions filtered by the SMACT enumeration algorithm. B is the formulas selected by BERTOS. $S \& B$ is the common part in S and B. $S-B$ is materials in S, but not in B. $B-S$ is materials in the B, but not in S.}
  \label{fig:energydist}
\end{figure*}

\paragraph{BERTOS for screening hypothetical materials compositions:}
Previously we only use Pymatgen's $oxid\_state\_guess$ or the enumerative algorithm of the SMACT package \cite{davies2018materials} for screening charge-neutral compositions generated by the deep generative models \cite{wei2022crystal}. However, one drawback of these oxidation state check methods is that they only consider mathematically possible combinations of accessible oxidation states of the constituent elements and neglect their preferred oxidation states in a given composition. We use our BERTOS model for large-scale hypothetical material screening. We first use the BLMM generative model \cite{wei2022crystal} to generate 1 million hypothetical compositions. After duplicate removal compared to the OQMD\cite{saal2013materials}, ICSD, and Materials Project \cite{jain2013commentary} databases, there are 635,064 new compositions left. We then filter out the binary and ternary compositions with the no. of atoms less than or equal to 30, leading to 175,798 compositions. Next, we apply two approaches to screen out charge-neutral compositions. First, as a baseline, we use the exhaustive enumeration algorithm for charge-neutrality checking as implemented in the SMACT package to filter out 119,221 compounds. Similarly, we apply our BERTOS to predict the oxidation states of the 175,798 compositions, pick up 64,046 hypothetical candidates, and compute the intersections and differences with those filtered by SMACT. 
We then predict the formation energies using the Roost model \cite{goodall2020predicting} (See Method section for details) for the compositions screened by our BERTOS and by SMACT, and plot their distributions as shown in the violin plots of Figure \ref{fig:energydist}. First, we find that compositions selected by SMACT ($S$) are located more at the higher energy region compared to those by BERTOS ($B$). There are 39,781 shared compositions ($S \& B$) by these two models, and the shape of $S \& B$ in the Figure \ref{fig:energydist} is very similar to the shape of $B$, which tend to have lower formation energy. We also show the distributions of those compositions are only appear in the SMACT set but not in BERTOS set (S-B) and compare it to those that only appear in the BERTOS set but not in the SMACT set (B-S). We find that SMACT tends to get more samples located at the higher end of the energy band, while the 6,266 unique samples selected by BERTOS tend to be located evenly toward the lower-energy area. These results show that BERTOS provides more stringent criteria and can select higher quality hypothetical material candidates. 
With high quality hypothetical compositions, one can use the modern template-based crystal structure prediction algorithms to get their structures \cite{wei2022tcsp,kusaba2022crystal} and use DFT calculations for further validation.

\paragraph{Discussion}
Given a material's structure, the chemical theories based bond valence sum method, structure descriptors based machine learning methods, and even DFT calculations may be used to obtain the actual oxidation states or charges of atomic sites. However, in computational materials discovery of novel materials, the structure is not even known making all existing methods not usable. Here we propose the first  only composition-based oxidation state prediction model for all element types based on the deep learning transformer language model neural networks, which is built for dealing with token sequence information with variable lengths. We formulate this oxidation state prediction problem as the token classification problem in natural language processing and surprisingly find that despite that our encoding of the compositions does not take any atomic properties into account except the elemental symbols themselves encoded with one-hot vectors, our transformer neural networks achieve superb performance for oxidation state prediction. Its OS prediction performance variations and difficulties are strongly consistent with known chemical knowledge: atoms with more flexible oxidation states are more difficult to predict their OS, which is especially true for non-metal elements. However, overall, the metal atoms are much more difficult to predict their OS compared to non-metal ones given the same number of possible oxidation states. Another unique advantage of our transformer based OS prediction is that all its predictions come with uncertainty quantification without using those tedious ensemble model based uncertainty estimation method \cite{varivoda2022materials}. The confidence score of  probability assigned to each atomic site OS makes it easy for users to interpret the prediction results. 

One of the important factors that affects the OS prediction performance is the amount of the training set and the distributions of the training and test sets. To investigate this issue, we trained four different BERTOS models using the four training datasets (See supplementary Table S1) including OS-ICSD, OS-ICSD-CN, OS-ICSD-oxide, and OS-ICSD-CN-oxide and test their performance over four different test sets (See supplementary file for detailed data processing steps to ensure there are no overlaps between the training sets and the test sets). Table \ref{tab:dataset_preformance} shows the OS prediction performance of different models over different test sets. First, over the OS-ICSD test set, the model trained with OS-ICSD training set achieves the best performance of 96.82\%. Since this test set has the most diverse compositions which include those that are not assigned with non-charge-neutral oxidation states, it requires the training set also have diverse samples. For the OS-ICSD-CN test set which contains only samples with charge-neutral oxidation state assignments by ICSD, the model trained with OS-ICSD achieves the best performance of 96.28\%, and the model trained with tOS-ICSD-CN has similar accuracy 96.27\%. For the OS-ICSD-oxide test set with only oxide samples, the OS-ICSD-oxide trained model achieves the best performance of 97.61\% as expected since OS-ICSD-oxide training set and test set have a more similar data distribution.
Finally, for the charge-neutral oxide test samples in the OS-ICSD-CN-oxide test set, the model trained with the OS-ICSD-oxide dataset achieves the best performance of 97.14\%, slightly beats the performance of 96.97\% of the model trained only with charge-neutral oxides in the OS-ICSD-CN-oxide training set with the similar argument for OS-ICSD-oxide test set. The reason may be that the OS-ICSD-oxide contains much more training samples (30,519) in addition to those oxides (20,601) in the OS-ICSD-oxide. Those non-CN training samples can be used to against the overfitting of the model, then can get the better performance. Overall, we find that the OS prediction performance of BERTOS for a given test set is the best or close to the best when the model is trained with a training set with similar compositions, indicating that more specialized training sets can produce higher performance for specialized test sets. However, increasing the training sets with additional diverse samples may also increase the OS prediction performance, especially when the training sets are relatively small.

\begin{table*}[th!]
\caption{OS prediction performance versus datasets}
\label{tab:dataset_preformance}
\begin{tabular}{
>{\columncolor[HTML]{FFFFFF}}c |
>{\columncolor[HTML]{FFFFFF}}c 
>{\columncolor[HTML]{FFFFFF}}c 
>{\columncolor[HTML]{FFFFFF}}c 
>{\columncolor[HTML]{FFFFFF}}c }
\hline
{\color[HTML]{000000} Train\textbackslash{}Test} & {\color[HTML]{000000} OS-ICSD} & {\color[HTML]{000000} OS-ICSD-CN} & {\color[HTML]{000000} OS-ICSD-oxide} & OS-ICSD-CN-oxide \\ \hline
OS-ICSD                                          & \textbf{96.82\%}               & \textbf{96.28\%}                           & 97.51\%                     & 97.11\%          \\
OS-ICSD-CN                                       & 95.92\%                        & \textbf{96.27\%}                  & 96.60\%                              &96.95\% \\
OS-ICSD-oxide                                    & 95.78\%                        & 94.96\%                           & \textbf{97.61\%}                     & \textbf{97.14\%}          \\
OS-ICSD-CN-oxide                                 & 94.95\%                        & 94.85\%                           & 96.70\%                              & 96.97\%          \\ \hline
\end{tabular}
\end{table*}

There are several ways our knowledge-agnostic deep learning model may be improved. For example, it may be useful to include the valence electrons and orbits information of the elements and  their common oxidation states into the element encoding. We may also conduct post-processing to fix the model prediction errors using known chemical rules such as charge-neutrality. We recognize that the actual oxidation states of atoms are also strongly dependent on its chemical environment so that the structure information is also important. However, it is surprising that our BERTOS achieves such high accuracy without the structure information. 

\section{Conclusion}
We proposed a transformer language model (BERT) based deep neural network, BERTOS, for composition based oxidation state prediction of inorganic materials. Extensive experiments have shown the unexpected high performance of our BERTOS models for assigning oxidation states to both metal and non-metal elements given only the composition/formula of a material. The model's performances over different families of elements and materials are all consistent with known chemistry knowledge. 
The high correlation between the number of accessible oxidation states of non-metals and the oxidation prediction performance of our models is interesting, as we did not observe such a strong correlation for metal elements. Since our model did not take into account any atomic attributes such as the valence electrons and orbital information, the self-emergence of the chemical knowledge learned by the knowledge-agnostic deep learning language models illustrates their big potential to explore new chemical knowledge. We still wonder and expect that when the elemental valence electron information and the orbital attributes are introduced to our model, the performance may be further improved or at least improved for a small training dataset. Compared to previous structure based oxidation state prediction models, our composition based oxidation state prediction models will have a large potential for screening millions of hypothetical new material compositions from generative design \cite{wei2022crystal,fu2022materials}.

\section{Methods}

Our raw dataset is obtained from the inorganic crystal structures database (ICSD)\cite{zagorac2019recent}. Each structural cif file of ICSD database contains the composition and corresponding oxidation states of each atomic site. We first remove those formulas with fractional OS or with more than 200 atoms and then remove all compounds to which only zero oxidation states are assigned (e.g. for most intermetallics). Since ICSD contains many CIF structure files that neglect the hydrogen atoms, we develop an algorithm to add the hydrogen atoms back to the structure along with their OS. To deal with the polymorphic phases issue (one composition has multiple crystal phases), we keep only one composition with the most frequent oxidation assignment. If several OS assignments have equal frequency, we pick the one with the maximum common oxidation states. 
In total, we obtain our OS-ICSD dataset with 52,147 samples. We further extract only charge-neutral compositions, which forms our OS-ICSD-CN dataset (37,424 samples). Similarly we obtain our OS-ICSD-oxide dataset (35,886 samples) by picking only oxides from the OS-ICSD dataset. Then we get the OS-ICSD-CN-oxide data set (24,229 samples) from the OS-ICSD-CN dataset by selecting formulas with the oxides. 
For each of these four datasets, we partition them into three subsets: training, validation, and test sets. The sample statistics are shown in Supplementary Table S1. Using this data partition approach, we guarantee that there is no overlaps between any pair of training set and test set. (See supplementary file for more details). 

\paragraph{Roost model training}
The Roost neural network model \cite{goodall2020predicting} for formation energy per atom prediction of candidate formulas is trained with 100,133 non-duplicate formulas from Materials Project, each formula picking their lowest formation energy per atom. We used the default network parameters with batch size 4096. The MAE error for the validation set is 0.108 eV. This model is one of the best models for composition based formation energy prediction.

\section{Code Availability}

The web app for the BERTOS can be tried at 
\hyperlink{http://www.materialsatlas.org/bertos}{http://www.materialsatlas.org/bertos}. The source code is freely available at \url{http://www.github.com/usccolumbia/BERTOS}. 

\section{Data Availability}
The pretrained models and datasets are freely available at \url{http://www.github.com/usccolumbia/BERTOS}.

\paragraph{Author contributions:}
Conceptualization, J.Hu.; methodology, N.F., J.H., Y.F., G.M., J.Hu; software, N.F., J.H., Y.F., J.Hu; resources, J.Hu.; writing--original draft preparation, J.H., N.F., Y.F., J.Hu., H.Z.; writing--review and editing, J.Hu, N.F. ; visualization, N.F., J.H., J.Hu; supervision, J.Hu.;  funding acquisition, J.Hu.
\paragraph{Competing interests: The authors declare that they have no competing interests}

\begin{acknowledgement}

Research reported in this work was supported in part by NSF under grants 1940099 and 1905775. The views, perspective, and content do not necessarily represent the official views of NSF.

\end{acknowledgement}

\begin{suppinfo}

Supplementary info can be found at \url{http://www.github.com/usccolumbia/BERTOS}

\end{suppinfo}

\bibliography{references}

\end{singlespace}

\end{document}